\input{aipcheck}

\documentclass[
    ,final            
  ]
  {aipproc}

\layoutstyle{6x9}

\begin{document}

\title{Fast X-ray Oscillations During Magnetar Flares}

\classification{97.10.Sj, 97.60.Jd, 95.85.Nv}
\keywords{neutron stars, torsional oscillations, pulsars, rotation, 
magnetospheres}

\author{Tod E. Strohmayer}{
  address={Astrophysics Science Division, NASA's Goddard Space Flight
Center, Greenbelt, MD} }

%

\begin{abstract}

The giant flares produced by highly magnetized neutron stars,
"magnetars," are the brightest sources of high energy radiation
outside our solar system. Serendipitous observations with NASA's Rossi
X-ray Timing Explorer (RXTE) of the two most recent flares resulted in
the discovery of high frequency oscillations in their X-ray fluxes.
The frequencies of these oscillations range from $\approx 20$ Hz to as
high as 1800 Hz, and may represent the first detection of global
oscillation modes of neutron stars. Here I will present an
observational and theoretical overview of these oscillations and
discuss how they might allow us to probe neutron star interiors and
dense matter physics.

\end{abstract}

\maketitle


\section{Introduction}

The existence of young ($< 10^4$ yrs) neutron stars with magnetic
fields upwards of $10^{15}$ G, ``magnetars,'' is now widely
accepted. Historically, two classes of neutron stars had been
associated with such strong magnetic fields, the Soft gamma-ray
repeaters (SGRs), and the Anomalous X-ray Pulsars (AXPs). The former
were first recognized as sources of super-Eddington, soft gamma-ray
bursts, while the latter were revealed as slowly spinning, rapidly
braking pulsars emitting much more X-ray flux than could be accounted
for by their spin-down energy alone.  In each case the energy
associated with the phenomena was attributed to decay of a
super-strong magnetic field.  In recent years the distinctions between
the two classes have steadily narrowed; for example, SGRs are now also
known to be ``anomalous'' pulsars, and AXPs have been observed to
burst (see Woods \& Thompson 2006 for a recent review of magnetars).

One distinction that remains is that the SGRs are, so far, the only
known sources of giant flares (also known as hyper-flares).  These are
the brightest cosmic events originating outside the solar system, in
terms of flux received at Earth. They are characterized by an intense
gamma-ray spike lasting tenths of seconds, and reaching luminosities
of $10^{44-46}$ ergs s$^{-1}$. Only three have been observed to date,
with the first being the famous March 5, 1979 gamma-ray flare from SGR
0526-66 in the LMC (Mazets et al. 1979). The two most recent giant
flares; from SGR 1900+14 in 1998 August (Hurley et al. 1999), and SGR
1806-20 in 2004 December (Palmer et al. 2005), are the subject of this
work.

Recent high time resolution studies of these flares using data from
the Proportional Counter Array (PCA) onboard the Rossi X-ray Timing
Explorer (RXTE) have resulted in the discovery of a new phenomenon
associated with these events. Both flares produced fast,
rotation-phase-dependent X-ray oscillations in the 20 - 150 Hz range,
and the 2004 flare additionally produced kHz oscillations in the 625 -
1,800 Hz range. Israel et al. (2005) first reported the discovery of
$\approx 18$, 30, and 90 Hz quasi-periodic oscillations (QPO) in the
December, 2004 event, and suggested that the 30 and 90 Hz QPOs could
be linked with seismic (torsional) vibrations of the neutron star
crust. Strohmayer \& Watts (2005) then reported the discovery of a
sequence of QPOs in the SGR 1900+14 event.  They found a set of
frequencies; 28, 53.5, 84, and 155 Hz, that could consistently be
associated with a sequence of low $l$ toroidal modes (denoted $_l
t_0$) of the elastic neutron star crust (see, for example, Hansen \&
Cioffi 1980; McDermott, van Horn \& Hansen 1988; Duncan 1998; Piro
2005).  In both flares the oscillations are episodic, that is, their
amplitudes vary considerably with time and rotational phase.  Watts \&
Strohmayer (2006) also examined Ramaty High Energy Solar Spectroscopic
Imager (RHESSI) data from the SGR 1806-20 event, confirmed the
presence of the 18 Hz and 90 Hz QPOs, and found evidence for
additional oscillations at 26 Hz and 626 Hz.  Most recently,
Strohmayer \& Watts (2006) reexamined the RXTE data from the SGR
1806-20 event, and found additional oscillations at 150, 625, and
1,840 Hz.  Figure 1 shows examples of QPO detections in each object.

\begin{figure}
\begin{tabular}{lr}
\includegraphics[width=2.95in, height=2.3in]{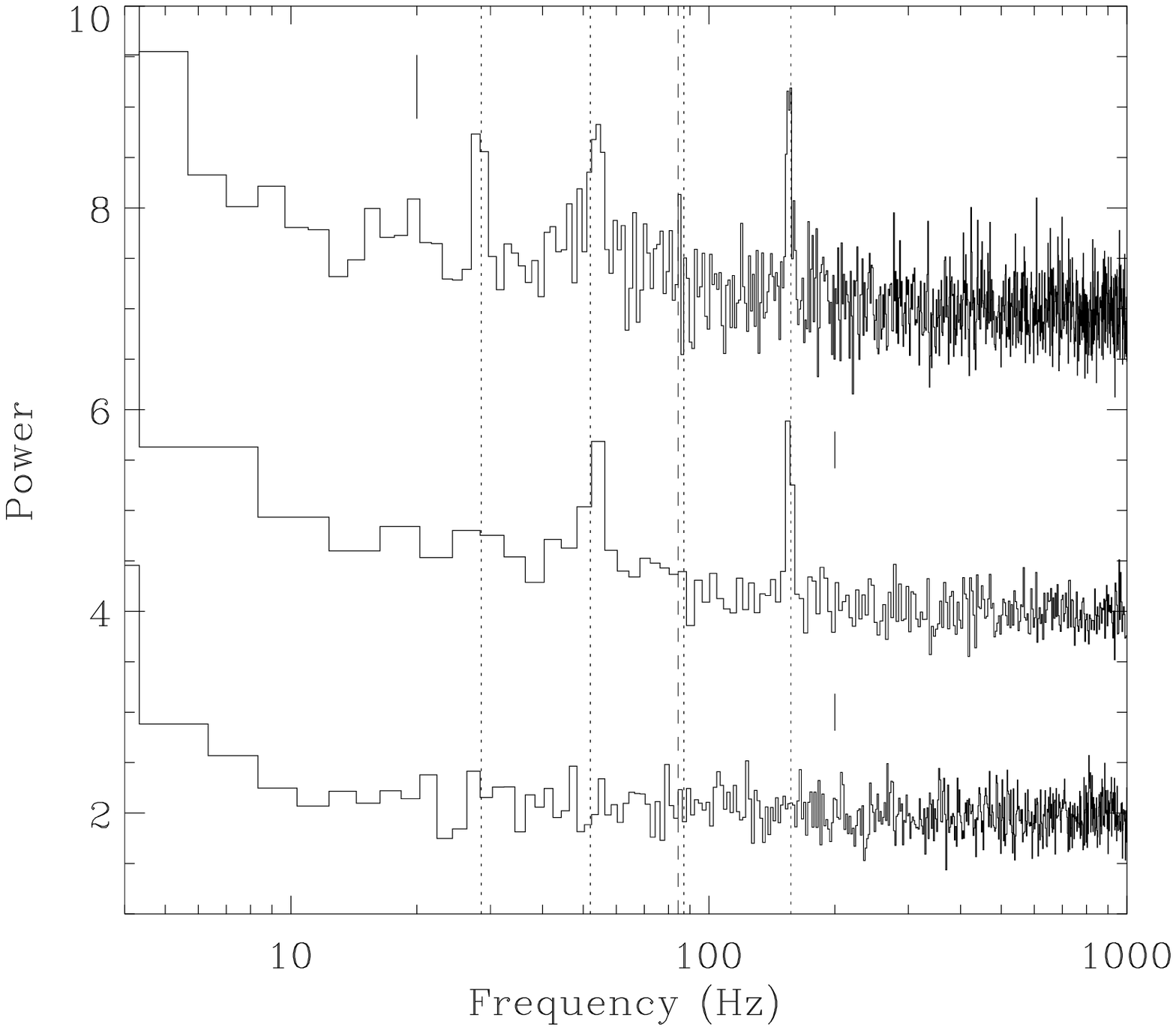}&
\includegraphics[width=2.95in, height=2.3in]{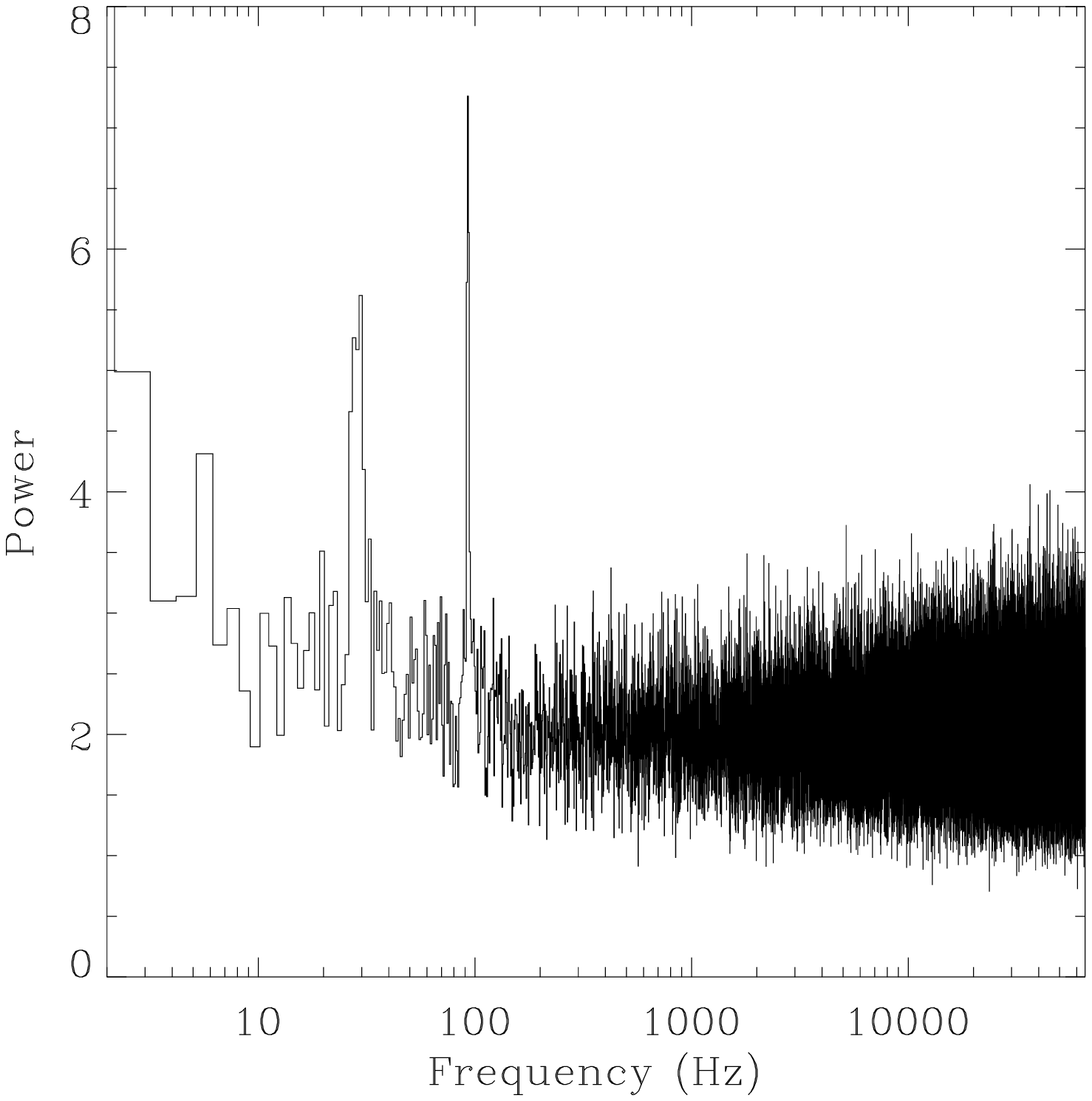} \\
\end{tabular}
\caption{Examples of QPOs detected during the magnetar flares from 
SGR 1900+14 (left) and SGR 1806-20 (right).  QPOs in SGR 1900+14 are
evident in an average power spectrum computed from rotational phases
centered on the phase at which an 84 Hz QPO was first detected (upper
traces), while no QPOs are seen at other phases (bottom trace).  QPOs
at 30 and 92 Hz are strongly detected in this average power spectrum
from the interpulse phase in the SGR 1806-20 flare (right). After
Strohmayer \& Watts (2005, 2006), respectively.}
\end{figure}

\section{Key Properties of the QPOs}

The similar phenomenology of the oscillations in the two sources, as
well as the closeness of some of the measured frequencies argues
convincingly that we are seeing the same physical process in each
case.  The connection with torsional modes of the crust seems
plausible for several reasons; 1) The observed frequencies are
consistent with theoretical expectations for such modes, and can be
more or less self-consistently associated with a sequence of modes
with varying spherical harmonic index, $l$. The higher frequency QPOs
(above 600 Hz) can plausibly be interpreted as modes with at least one
node in the radial displacement eigenfunction. 2) The magnetic
instability which powers the flares will very likely fracture the
neutron star crust, generating seismic motions within the star
(Flowers \& Ruderman 1977; Thompson \& Duncan 1995; Duncan 1998;
Thomson \& Duncan 2001; Schwartz et al. 2005). 3) The strong
rotational phase dependence argues for a mechanism associated with
particular sites on the stellar surface, such as a fracture zone or
magnetic field bundle. 4) Mechanical motions provide a natural
explanation for the relatively high coherence of the oscillations.  5)
Horizontal motions of the crust could modulate the beaming pattern
associated with the strong magnetic field, providing a mechanism to
modulate the X-ray flux. Moreover, beaming can act as an
``amplifier,'' producing potentially large X-ray modulations from
modest horizontal displacements.  Although the present evidence for
torsional modes is very suggestive, it is not yet definitive. Levin
(2006), for example, has argued that toroidal modes may damp too
quickly to account for the detection of oscillations some minutes
after the onset of the flare.  In addition some of the detected
frequencies (18 Hz, 26 Hz) do not fit easily into current torsional
mode models, without invoking magnetic splitting or other
complications, such as coupling of the crust with the core.

\section{Torsional Mode Interpretation}

Since the detections of the magnetar QPOs there has been a substantial
theoretical effort to interpret the observed frequencies, amplitudes
and other properties. As mentioned above, the initial suggestions
focused on oscillations of the neutron star crust, the so called
torsional modes, denoted $_l t_n$, where $l$ is the spherical harmonic
index for the mode, and $n$ is the number of radial nodes in the
eigenfunction (Israel et al. 2005; McDermott, Van Horn \& Hansen
1988).  These modes sense the shear wave speed in the crust, and its
size (see, for example, Strohmayer et al. 1991; Duncan 1998). These
quantities depend on the global structure of the star and hence the
equation of state (EOS) of matter in the deep interior (see Strohmayer
\& Watts 2005; Lattimer \& Prakash 2007). Note also that the strong 
magnetic field can boost the ``tension'' in the crust and thus modify
the mode periods, perhaps non-isotropically (see Duncan 1998; Messios,
Papadopolous \& Stergioulas 2001). Recently, Samuelsson \& Andersson
(2007) have explored torsional mode oscillations in neutron stars
using a general relativistic formulation with neglect of the metric
perturbations (the so called Cowling approximation). They derive a set
of analytic estimates for the mode frequencies based on numerical
calculations, and estimate the frequency (in Hz) of the $n=0$ modes to
be
\begin{equation}
f(_l t_0) = 27.65 \; R_{10}^{-1} \; \frac{\left ( (l-1)(l+2)\right
 )^{1/2} }{2} \frac{\left ( (1.705 - 0.705\beta_{*})(0.1055 +
 0.8945\beta_* ) \beta_* \right )^{1/2}}{(1.0331\beta_* - 0.0331)} \; ,
\label{e1}
\end{equation}
where $R_{10} = R/10$ km and $\beta_* = \beta / 0.2068$, and $\beta =
GM/c^2 R$ (for $R_{10} = 1$ and $M = 1.4 M_{\odot}$, $\beta_* =
1$). The expression for the $n > 0$ modes (again, in Hz) is
\begin{equation}
f(_l t_n) = 1107.3 \; \frac{n(0.1055 +
0.84945\beta_*)}{R_{10}}\frac{\beta_*}{1.0166\beta_* - 0.0166} \; .
\label{e2}
\end{equation}
These expressions do not include any correction for magnetic field
effects. Note that the scaling with $l$ here is slightly different
than the $l(l+1)$ scaling in previous estimates (see Duncan 1998, Piro
2005).  This can lead to somewhat different mode assignments for
particular observed frequencies. Since these relations depend on both
the stellar mass and radius they are clearly EOS dependent, and thus
secure mode identifications could provide constraints on the
EOS. However, the effects of the magnetic field on the oscillation
mode spectrum will likely have to be understood quantitatively before
precise EOS constraints will be possible.  Nevertheless, the promise
of using asteroseismology to constrain the neutron star EOS is
undeniable.

\begin{figure}
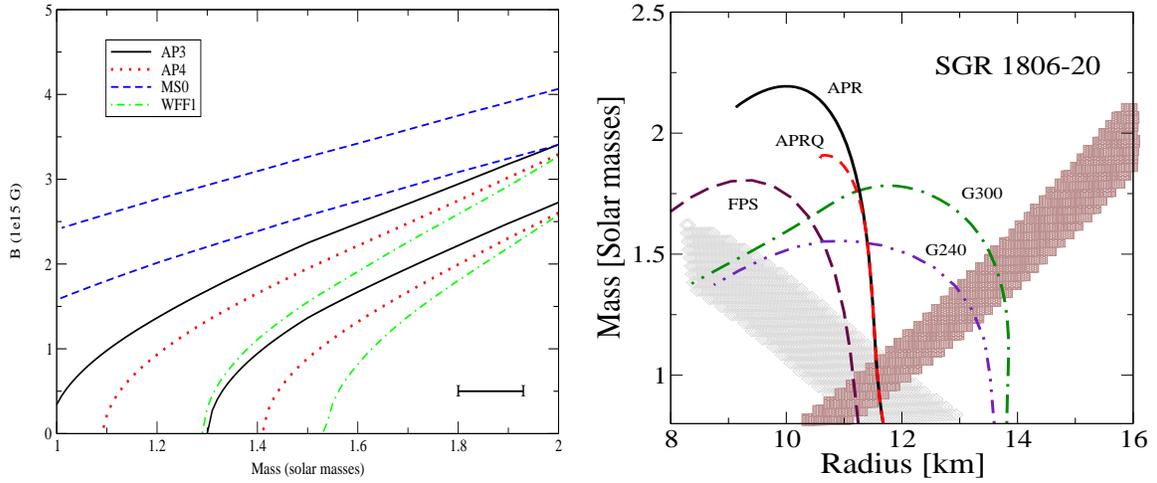

\begin{tabular}{lr}
\includegraphics[width=2.9in, height=2.5in]{f2a.eps}&
\includegraphics[width=2.9in, height=2.5in]{f2b.eps} \\
\end{tabular}
\caption{Constraints on neutron star structure from magnetar QPOs. The 
stellar mass and magnetic field required to give the $_2 t_0$ mode
frequencies inferred for SGR 1806-20 (30.4 Hz), and SGR 1900+14 (28
Hz) are shown in the left panel (after Strohmayer \& Watts 2005) for
several different EOS models. For each EOS, two lines are shown. The
upper line is for SGR 1806-20, and the lower line for SGR 1900+14. The
regions allowed based on $n=0$ and $n=1$ mode identifications in SGR
1806-20 of Samuelsson \& Andersson (2007) are shown in the right panel
(after Samuelsson \& Andersson 2007). The shaded region with
decreasing mass versus radius denotes the $n=0$ constraints.  The
orthogonal nature of the constraints for $n=0$ and $n>0$ modes clearly
demonstrates the potential power of detecting both types of modes.}
\end{figure}

To illustrate the possibilities we review two examples from the recent
literature. Strohmayer \& Watts (2005) tabulated the stellar
parameters that give $_2t_0$ oscillations at 28 Hz (SGR 1900+14) and
30.4 Hz (SGR 1806-20).  These were the fundamental mode
identifications that they made based on the mode periods, including an
isotropic magnetic correction, estimated by Duncan (1998). Figure 2
(left) shows the results for four different EOS discussed in Lattimer
\& Prakash (2001). The results suggest that if the stars have similar
magnetic field strengths, their masses must differ by more than $0.2
M_\odot$. Since the masses of radio pulsars (a young neutron star
population) have been found to be consistent with a fairly narrow
Gaussian distribution, $M=1.35\pm 0.04 M_\odot$, a perhaps more likely
scenario is that the stars have similar masses but different magnetic
field strengths.  If both stars have masses $\approx 1.35 M_\odot$,
then it is difficult to account for both frequencies with the softest
EOS (WFF1); and the stiffest EOS (MS0) predicts magnetic fields for
both systems that are far larger than those inferred from timing
studies (Woods et al. 2002).  The moderately stiff EOSs AP3 and AP4
can account for the observed frequencies, and give magnetic field
strengths that agree reasonably well with those derived from timing
measurements of both stars (Woods \& Thompson 2006).  Samuelsson \&
Andersson (2007) used their relativistic formulation to determine the
range of stellar masses and radii which could produce acceptable
matches to the observed frequencies for some sequence of $l$
values. Their results for SGR 1806-20 are also shown in Figure 2
(right). They are able to associate the $\approx 29$, 93, and 150 Hz
QPOs with $l = 2, 6$ and 10 modes (all with $n=0$). The higher
frequency QPO at 626 Hz is associated with an $n=1$ mode for which $l$
is not greatly constrained.  The allowed regions for the $n=0$ and
$n=1$ modes are largely orthogonal, and in principle, could provide
very tight constraints on the stellar parameters.  

\section{Remaining Theoretical Challenges}

\subsection{Crust - Core Coupling}

The question of whether or not pure crust models are adequate is an
important one. Levin (2006) argued that if there is a strong
perpendicular magnetic field threading the crust - core boundary, then
large horizontal displacements at the base of the crust must excite
vibrations in the core, perhaps significantly damping crust motions.
The effects of coupling may, however, be mitigated in several
ways. For example, coupling to the core will depend sensitively on the
amplitude at the boundary, and previous calculations show that this
amplitude can be substantially smaller than that at the top of the
crust (see McDermott et al. 1988, for example). The physics at the
boundary layer at the base of the crust will also be crucial (see, for
example, Kinney \& Mendell 2003). The elastic properties at the base
of the crust (Pethick \& Potekhin 1998) have not yet been considered,
and could substantially modify the eigenfunctions in this region. In
addition, coupling will likely depend rather sensitively on both the
magnetic field geometry and the particular displacement pattern of
individual modes. The modes that persist, and that we observe, may be
those for which the coupling is minimal. The presence of a strong
toroidal field in the core of the star could also reduce coupling by
making the core more rigid and less prone to excitation.

If the coupling between crust and core is strong, by virtue of
boundary layer physics or magnetic field threading, one needs to
consider the global magneto-elastic modes of the neutron star.  As
first suggested by Israel et al. (2005), global modes can accommodate
the lower frequency 18 and 26 Hz QPOs very easily.  A recent paper by
Glampedakis et al. (2006) developed a simple slab model of global
magneto-elastic oscillations that showed two interesting features.
Firstly, it confirmed the presence of modes at lower frequencies than
the ``pure crust'' toroidal modes.  Secondly, the model exhibited
modes for which the amplitudes in the crust were strong; in these
cases the frequency was very close to the well-established ``pure
crust'' frequencies. In other words, even with coupling included, it
is possible to obtain very similar frequencies to those that we know
match the data. Although the model was very simple (slab geometry, for
example, is not adequate to describe behavior deep in the core), this
gives us some indication that global modes may have similar
frequencies.

\begin{figure}
\includegraphics[width=4.0in, height=3.25in]{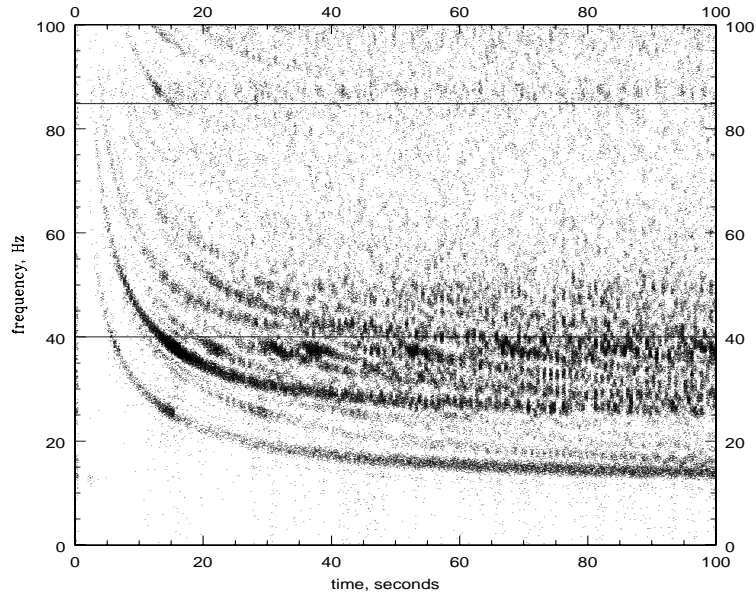}
\caption{A dynamic power spectrum computed from a physical simulation 
of the crustal displacement in a magnetar, including the interaction
between crust and core, is shown (after Levin 2007). The density of
points tracks the Fourier power. The low frequency QPOs asymptotically
approach the MHD core continuum turning points (see Levin 2007 for a
detailed discussion). Interestingly, excess oscillation power also
appears near the frequencies of crustal torsional modes (horizontal
lines).  }
\end{figure}

In a more recent calculation, Levin (2007) has explored in some detail
the coupling between normal modes in the crust and an MHD continuum in
the fluid core, in the context of a uniform magnetic field and thin
spherical crust. He finds that the torsional modes in the crust
quickly exchange energy with the core MHD continuum, and that there
are natural QPOs associated with the dynamics of such a system.  The
QPOs are associated with either the ``edges'' of the continuum, or at
so called ``turning points,'' where there is a local extremum in the
frequency distribution of the continuum.  Levin describes a relatively
simple, but physically well motivated toy model in which a large
number of small pendula are coupled to a single large pendulum.  The
frequency of the large pendulum lies in the middle of the range of the
smaller ones. In this simulation the big pendulum represents an
initially stressed magnetar crust, and the small pendula represent the
MHD core continuum with which it is coupled. After release, the motion
of the big pedulum is rapidly damped, as energy is drained into the
smaller pendula.  In this system, Levin finds QPOs at the edges of the
continuum, that is, at the lowest and highest frequencies of the
little pendula. The QPOs occur at these frequencies, because it is
only in the vicinity of these points where the pulling of individual
small pendula on the large one are not cancelled out by other small
pendula. While the real magnetar dynamics are much more complex, Levin
argues, rather convincingly, that this simple model captures much of
the relevant physics.  In a more realistic simulation of the crust -
core coupling he finds interesting qualitative agreement with many of
the observed properties of the magnetar flares.  In particular, strong
QPOs at about 18 Hz are present and would appear to be consistent with
the turning point of the MHD core continuum. We show in Figure 3 the
dynamic power spectrum from Levin's calculation, and the QPOs are
clearly evident as the darker bands in the spectrum, particularly at
the lower frequencies (18 - 30 Hz).  Moreover, QPOs appear to be
excited or amplified at frequencies near the pure crustal mode
frequencies. Interestingly, the QPOs in these simulations also show a
clear drifting in frequency with time which remains to be clearly
understood.  While this initial theoretical work appears very
promising, there are still many remaining questions, and more
realistic physical scenarios need to be addressed, including more
realistic magnetic field geometries. More work in this area is clearly
warranted.

\subsection{Magnetospheric Coupling: X-ray Modulation}

While the evidence is now substantial that the magnetar oscillations
are due to neutron star vibrations, there is still a great deal of
uncertainty on the details of how stellar surface motions get
translated into rather strong modulations in the observed X-ray flux.
Moreover, the present data hint at a complex temporal evolution of the
oscillations. For example, some of the oscillations are seen
throughout the flare, others are detected only half way through the
tail, and in general the highest frequency modes appear to be rather
short-lived.  This, and other, complexities require explanation, and
to do so we must understand the physics of mode excitation, damping,
and X-ray modulation.

It has been suggested that the ``shaking'' of the magnetic field lines
due to stellar vibrations, coupled with strong beaming of the emitted
radiation, could play an important role in the modulation mechanism
(see Strohmayer \& Watts 2006).  Recently, Timokhin, Eichler \&
Lyubarsky (2007) have proposed that modulation of the X-ray flux is
linked to variations of the magnetospheric currents induced by
oscillatory motions of the surface (via the magnetic field). They show
that the angular distribution of the optical depth to resonant Compton
scattering produced by a particular torsional mode pattern can be
highly anisotropic, and argue that this can account for the observed
rotational phase dependence of the oscillations.  Moreover, they
estimate the amplitude of surface motions required to produce the
observed oscillation amplitudes, and find that this is about 1\% of
the stellar radius, or about 100 meters.  This amplitude represents a
substantial amount of mechanical energy, Timokhin et al. (2007)
estimate $\approx 10^{42}$ ergs per mode would need to be
deposited in crustal motions by the giant flares. A significant amount
of this energy would likely end up as heat in the crust, and could
power some of the long term X-ray afterglows that have been observed
(Kouveliotou et al. 2003).  A nice feature of this model is that it
fits well with an emerging consensus that the hard X-ray emission from
magnetars is associated with scattering by magnetospheric currents
(Lyutikov \& Gavriil 2006; Fern{\'a}ndez \& Thompson 2007).  However,
more realistic, quantitative predictions will be required to make
detailed comparisons with observations.

\section{Conclusions}

Serendipitous RXTE observations of the magnetar flares from SGR
1806-20, and SGR 1900+14 indicate that a complex pulsation
phenomenology is associated with these events.  The discovery of new
kHz-range frequencies consistent with theoretical predictions for
$n>0$ torsional modes provides strong evidence that we may in fact be
seeing vibration modes of the neutron star crust excited by these
catastrophic events. If this is true, then it opens up the exciting
prospect of probing the interiors of neutron stars in a manner
analogous to helioseismology.  Additional excitement is warranted when
we consider that all the current datasets used to explore these
oscillations have been purely serendipitous.  That is, they have not
been optimized in any way for studying these signals.  This suggests
that a wealth of additional information would likely be found from
instruments better optimized to capture with high time resolution the
flood of X-rays produced by these events.  In order to fully exploit
such observations more theoretical work is definitely needed to enable
more accurate mode identifications, to better understand the
excitation and damping mechanisms of modes and how they can couple to
the X-ray emission, and to make more precise inferences on neutron
star structure.





\bibliographystyle{aipprocl} 



\end{document}